# Neutron capture by $^{15}$N nucleus at astrophysical energies

S B Dubovichenko, N V Afanasyeva

Fessenkov Astrophysical institute "NCSRT" NCA RK, 050020, Almaty, Kazakhstan

E-mail: dubovichenko@gmail.com, n.v.afanasyeva@gmail.com

**Abstract.** Within the potential cluster model with the forbidden states and an orbital states classification according to the Young diagrams the possibility of description of experimental data for the total cross-sections of radiative n$^{15}$N-capture at energies from 25 to 370 keV was considered. It was shown that only on basis of the E1-transitions from the different states of n$^{15}$N-scattering to the ground state of $^{16}$N nucleus in n$^{15}$N-channel it is well succeed to explain the value of the total cross-sections in the considered energy range and to prognosticate its behavior at energy E ≤ 1 eV. These cross-sections at energies E ≤ 10 keV are approximated by the simple analytical form.

## 1. Introduction

Continuing the study of the reactions of neutrons radiative capture by light nuclei which take part in the various thermonuclear processes [1] let's discuss the reaction of n$^{14}$N→$^{15}$N$\gamma$ capture at low energies. This process is included in the main chain of primary nucleosynthesis reactions [2,3,4] which determined the development of the Universe in an early stages of its forming [3].

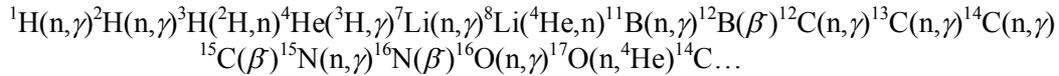

$^{1}$H(n,$\gamma$)$^{2}$H(n,$\gamma$)$^{3}$H($^{2}$H,n)$^{4}$He($^{3}$H,$\gamma$)$^{7}$Li(n,$\gamma$)$^{8}$Li($^{4}$He,n)$^{11}$B(n,$\gamma$)$^{12}$B($\beta$)$^{12}$C(n,$\gamma$)$^{13}$C(n,$\gamma$)$^{14}$C(n,$\gamma$)$^{15}$C($\beta$)$^{15}$N(n,$\gamma$)$^{16}$N($\beta$)$^{16}$O(n,$\gamma$)$^{17}$O(n,$^{4}$He)$^{14}$C…

Earlier in our works [5,6,7,8,9,10,11] the astrophysical S-factors and the total cross-sections of 20 reactions (including n$^{15}$N→$^{16}$N$\gamma$) of radiative capture of protons, neutrons and other particles by many light atomic nuclei were considered. For such processes analysis we usually use the methods of calculations based on the potential cluster model (PCM) of light atomic nuclei with the forbidden states (FS) [12,13]. The presence of the FS in the wave functions (WF) is defined on the basis of the orbital states classification according to the Young diagrams. This classification was described in works [13,14]. In the used approach the potentials of intercluster interactions for the scattering processes are constructed on the basis of reproduction of the considered particles elastic scattering phases taking into account the resonance behavior of these particles [15]. For the bound or ground (BS or GS) states of the nuclei generated in the reaction in the cluster channel which is the same as the initial particles, the intercluster potentials are constructed on the basis of the description of these particles binding energy in the final nucleus and some basic characteristics of such states [16]. For example, the root mean square radius of the final nucleus and its asymptotic constant (AC) in this channel are considered.

Meanwhile for any cluster system the many-particle character and antisymmetrization effects are qualitatively taken into account by such potential one-particle levels separation for the allowed (AS) and forbidden by the Pauli principle states [6-13,16,17]. Choice of the potential cluster model with the forbidden states for the consideration of such cluster systems is caused by that in many light atomic nuclei the probability of the nucleon associations, i.e. clusters, forming and their isolation degree are rather high. It is confirmed by many experimental data and various theoretical calculations obtained by different authors in the last 50-60 years.

## 2. Model and methods

Proceeding to n$^{15}$N-capture total cross-sections analysis let's note that $^{15}$N nucleus orbital states classification according to the Young diagrams was qualitatively considered by us in works [6,17]. So for n$^{15}$N system we have {1} × {4443} → {5443} + {4444}. The first of the obtained diagrams is compatible with the orbital moment $L$=1, 3 and forbidden as it has 5 boxes in the first row, and the second one is allowed and compatible with the orbital momentum $L$=0 [14]. Thus, taking into account only the lowest partial waves it may be said that in the potentials of $S$- and $D$-waves the forbidden states are absent, but the $P$-wave contains the forbidden state. The allowed in the D-wave state corresponds to the ground state of $^{16}$N nucleus at n$^{15}$N-system binding energy -2.491 MeV [18]. Since the moment of $^{15}$N nucleus is equal to $J^\pi T$ = 1/2$^-$1/2 [19] and $^{16}$N nucleus characteristics are known $J^\pi T$ = 2$^-$1 [18] then $^{16}$N ground state in n$^{15}$N-channel may be the mixture of $^1D_2$- and $^3D_2$ states in $^{2S+1}D_J$ notations.

Since we have not the total tables of the Young diagrams product for the particle systems with A>8 [20], which were used by us earlier for such calculations [5-11,21,22], therefore the obtained above result should be considered as only the qualitative evaluation of possible orbital symmetries for the bound states of $^{16}$N nucleus in n$^{15}$N-channel. At once just on the basis of such classification it was succeeded to explain well enough the available experimental data for the radiative p$^{13}$C- [23], n$^{13}$C [6,24] capture and also for n$^{14}$N [6,17] capture. So here we will use this given above classification of cluster states according to the orbital symmetries. This classification gives the certain number of the FS and AS in the different partial intercluster interaction potentials. Number of these states defines the number of nodes of the relative motion radial WF of the clusters with certain orbital moment [7].

The expressions for calculating of the radiative capture process $\sigma(NJ)$ total cross-sections in the cluster model have the following form (see, for example, [25] or [16,26,27,28]):

$$\sigma_c(NJ, J_f) = \frac{8\pi K e^2}{\hbar^2 q^3} \frac{\mu}{(2S_1+1)(2S_2+1)} \frac{J+1}{J[(2J+1)!!]^2} A_J^2(NJ,K) \sum_{L_i,J_i} P_J^2(NJ,J_f,J_i) I_J^2(J_f,J_i), \quad (1)$$

where for the orbital electric $EJ(L)$ transitions the following expression is known [25-28]:

$$P_J^2(EJ, J_f, J_i) = \delta_{S_i S_f} \left[ (2J+1)(2L_i+1)(2J_i+1)(2J_f+1) \right] (L_i 0 J 0 | L_f 0)^2 \begin{Bmatrix} L_i & S & J_i \\ J_f & J & L_f \end{Bmatrix}^2,$$

$$A_J(EJ,K) = K^J \mu^J \left( \frac{Z_1}{m_1^J} + (-1)^J \frac{Z_2}{m_2^J} \right), \quad I_J(J_f, J_i) = \langle \chi_f | R^J | \chi_i \rangle.$$

Here $S_1$, $S_2$, $L_f$, $L_i$, $J_f$, $J_i$ are the spins, orbital and total moment of particles in the initial channel (*i*) and the nucleus in the final channel (f) correspondingly; $\mu$, $q$ are the reduced mass and the wave number of particles in the initial channel; $m_1$, $m_2$, $Z_1$, $Z_2$ are the masses and charges of particles in the initial channel; $K^J$, $J$ are the wave number and the moment of $\gamma$ in the final channel; $I_J$ is the integral over the initial and final states wave functions, i.e. over the clusters relative motion functions with the intercluster distance $R$, $N$ is $E$ or $M$ transitions of multipolarity $J$ from the $J_i$ initial to $J_f$ final state of nucleus [26-28].

Let's emphasize that in our present or previous [7,16,26,28] calculations we always take the spectroscopic factor $S_f$ equal to 1. This spectroscopic factor may enter into the expression for the cross-section (see for example [25]). Apparently, considering the capture reaction in the potential cluster model it is not necessary to include the additional $S_f$ factor [6,7] for the potentials consistent in the continuous spectrum with the scattering processes characteristics taking into account the resonant form of phases and also for the discrete spectrum potentials describing the basic nucleus GS or BS characteristics. As the result all existing in the reaction effects expressed in the certain factors and coefficients are taking into account by the interaction potentials, because these potentials are constructed on the basis of the description of observable characteristics. These characteristics are the experimental characteristics of the clusters interacting in the initial channel and some generated in the final state nucleus described by the cluster structure consisting of the initial particles.

For the performing of the radiative capture total cross-sections calculations our computer program based on the finite-difference method was rewritten [27,28]. Besides the finite-difference method the variational method with the wave function expansion in an unorthogonal gauss basis with the independent variation of the expansion parameters [7,27,29] was used.

$$\Phi_L(R) = \frac{\chi_L(R)}{R} = R^L \sum_i C_i \exp(-\beta_i R^2),$$

where $\beta_i$ are the variational parameters and $C_i$ are the expansion coefficients. Variational program [27] was used in order to check the accuracy of the calculation of the binding energy and form of the considered nucleus wave function in the ground and bound states by using the finite-difference method.

In these calculations the exact value of neutron mass [30] with $^{15}$N nucleus mass equal to 15.000108 a.m.u. [31] was assigned, and the value of $\hbar^2/m_0$ constant was taken equal to 41.4686 MeV·Fm$^2$. Although this value is considered as rather obsolete to date in comparison with the new value 41.801591 MeV·Fm$^2$ [32], we continue to use it for facilitation of the comparison of the last [5-11] and all earlier obtained results [15,16,26].

**3. Interaction potentials**
For the radiative n$^{15}$N-capture total cross-sections description as well as in the previous works [6-11,15,17,22-24] we considered $E1$-transitions. In the given process this transition is possible from the nonresonance $^3P_2$ scattering wave with zero phases at energies up to 1.0 MeV to the triplet part of $^3D_2$ wave function of the bound ground state of n$^{15}$N-clusters in $^{16}$N-nucleus. In addition $E1$-transitions are possible from the resonance $P_1$ scattering wave at 0.921 MeV [18], which is the mixture of the $^3P_1$ triplet and $^1P_1$ singlet states, to the $^3D_2$- triplet and $^1D_2$ singlet part of the ground state wave function correspondingly. On the analogy with work [33] let's write the cross-section in the following form

$$\sigma_0(E1) = \sigma(E1, ^3P_2 \to ^3D_2) + \sigma(E1, ^3P_1 \to ^3D_2) + \sigma(E1, ^1P_1 \to ^1D_2).$$

For performing the calculations of the radiative capture total cross-sections the nuclear part of intercluster n$^{15}$N-interaction potential as usual is presented in the Gaussian form [5]

$$V(r) = -V_0 \exp(-\alpha r^2).$$

Let's notice that for the potential of resonance $^{1+3}P_1$ scattering waves at 0.921 MeV [18] with one FS it is not succeed to obtain the potential which is able to reproduce correctly the width of resonance. For example, the potential with parameters $V_{P1}$ = 7687.40 MeV, $\alpha_{P1}$ = 10.0 Fm$^{-2}$ leads to the resonance width equal to 138 keV (the center-of-mass system) at energy 0.921 MeV that is approximately in 10 times greater than experimental value 14 keV (the laboratory system) given in table 16.10 of work [18]. Although according to the data presented in table 16.5 of the same review [18] this width is equal to 15(5) keV in the center-of-mass system. Calculation of $P_1$-phase of elastic n$^{15}$N-scattering with such potential at energies from 0.2 to 1.25 MeV shows that it has the resonance form represented in figure 1 by solid curve. In order that this potential may correctly reproduce the width of resonance it would be necessary to increase greatly $\alpha$ parameter, i.e. to decrease width of potential. For example, the parameters 15385.47 MeV and 20.0 Fm$^{-2}$ lead to the width of resonance about 98 keV in the center-of-mass system. This potential phase is shown in figure 1 by dotted curve. Parameters of much more narrow potential with one FS are

$$V_{P1} = 30781.774 \text{ MeV}, \alpha_{P1} = 40.0 \text{ Fm}^{-2}. \qquad (2)$$

They lead to the resonance width 70 keV in the center-of-mass system and its phase is represented in figure 1 by dashed curve. Relative accuracy of $P_1$ scattering phase computing in these calculations is approximately $\pm 10^{-3}$, and this potential leads to the phase value 90.0(1)° at the resonance energy 921 keV.

As one can see from these calculations even for very approximate description of this resonance width in the elastic n$^{15}$N scattering the $P_1$-potential with absolutely exceptional width can be obtained. So for the independent checking of these results other program for the phases calculation was used. This program was independently developed by the outside producer and it is based on other computing methods for the wave function calculation and rather different ways of matching of the wave function with asymptotic [34]. On the basis of it the scattering phase equal to 89.9° was obtained for the potential (2) at resonance energy 921 keV (the laboratory system) and specifying of 500 thousand steps of the wave function calculation and matching of it with asymptotic at 30 Fm. This result differs from the other one given above only on 0.1°, i.e. on the phase estimation error evaluation specified in our program with automatic selection of step and the wave function matching radius. Thereby the results obtained by using two independent programs totally confirm the rightness of this potential construction. It is necessary to keep in mind that in all our calculations the $\hbar^2/m_0$ constant was equal to 41.4686 MeV·Fm$^2$.

Here it is necessary to notice that the potential with the FS at known energy of resonance level in $^{16}$N nucleus spectra and its width is constructed absolutely unambiguously. It is not possible to find other parameters $V_0$ and $\alpha$ which would be able to reproduce correctly the resonance level energy and its width, if the number of FS is given, which in this case is equal to 1. Such potential depth unambiguously locates a position of resonance, i.e. the energy of resonance level, and its width gives the definite width of this resonance state. However, we didn't succeed to find some physical interpretation of so small width and huge depth of this potential. The total mass of 16 nucleons in nucleus is approximately equal to 15 GeV, and the depth of potential (2) takes on a value 30 GeV. Thus, used here potentials should be considered as an attempt of rough approximation within the potential approach (i.e. on the basis of interaction potentials taking into account a resonance behavior of scattering phases of two nuclear particles with given masses) of experimentally observed width of the considered level at energy 921 keV which is present in elastic n$^{15}$N-scattering (see table 16.10 from review [18]).

For the potentials of nonresonance $^3P_2$- and $^3P_0$-waves with one FS we have used the parameters' values based on the assumption that in the considered range of energies, i.e. up to 1.0 MeV, the potentials phases are equal to zero, as the resonance levels with $J = 2^+$ and $0^+$ (see table 16.10 [18]) are not observed in $^{16}$N nucleus spectra in n$^{15}$N-channel. Particularly, it has been obtained that

$$V_{P2} = 500.0 \text{ MeV}, \alpha_{P2} = 1.0 \text{ Fm}^{-2}. \qquad (3)$$

Calculation of $P$-phases with such potential at energy up to 1.0 MeV leads to their values less than 0.1°.

The potential of the $^{1+3}D_2$-ground state without bound FS must correctly reproduce the binding energy of $^{16}$N nucleus ground state with $J^\pi, T = 2^-, 1$ in n$^{15}$N channel at 2.491 MeV [18]. Also this potential must reasonably describe a root mean square radius of $^{16}$N nucleus, experimental value of which evidently mustn't exceed a lot $^{16}$O nucleus radius equal to 2.710(15) Fm [18]. Note, that the experimental radius of $^{15}$N nucleus is equal to 2.612(9) Fm [19]. In these calculations we use a zero charge radius of neutron the mass radius of which is equal to proton radius 0.8775(51) Fm [30]. As a result the following parameters for the potential of $^{16}$N nucleus ground state in n$^{15}$N-channel were obtained

$$V_{g.s} = 49.5356532 \text{ MeV}, \quad \alpha_{g.s} = 0.07 \text{ Fm}^{-2} \qquad (4)$$

The potential leads to the binding energy -2.49100003 MeV at the finite-difference method accuracy $10^{-8}$ MeV, the root mean square radius 2.63 Fm and mass radius 2.76 Fm. For the AC defined in work [35] and written in the nondimensional form

$$\chi_L(R) = \sqrt{2k_0} \, C_0 W_{-\eta L+1/2}(2k_0 R)$$

the value 0.96(1) was obtained over the range 6÷19 Fm. Error of the constant is defined by its averaging over the mentioned above distance range. In work [36] this AC has a value 0.85 Fm$^{-1/2}$, that after the recalculation to the nondimensional quantity at $\sqrt{2k_0} = 0.821$ leads to value 1.04. This value differs from the other one obtained above only on 10%. The recalculation of the AC is needed because in work [36] other definition of the AC was used

$$\chi_L(R) = C W_{-\eta L+1/2}(2k_0 R),$$

that differs from the used here definition on a factor $\sqrt{2k_0}$.

The ground state potential is constructed relatively unambiguously because its depth depends on the number of the bound states allowed or forbidden in this partial wave, and the potential width is defined by the AC value.

For additional control of calculation of the ground state energy the variational method [27,28] was used. This method for the grid dimension $N=10$ and independent variation of parameters for potential (4) let to obtain the energy -2.49100001 MeV. The value of the AC is equal to 0.96(1) over the range 9÷24 Fm, and the charge radius doesn't differ from other one obtained above within the finite-difference method. As for the basis dimension increasing the variational energy decreases and gives the upper limit of true binding energy [7], and for the stride parameter decreasing and steps number increasing the finite-difference energy increases [6,7,27,28], so the real binding energy in such potential can take on the average value -2.49100002(1) MeV. Thus, the accuracy of calculation of two-body binding energy of the nucleus by using two methods (the finite-difference method and the variational method) and two different and independent computer programs is equal to $\pm 10^{-8}$ MeV = $\pm 10$ meV. This value is agree with the primordially given in the finite-difference method accuracy in case of calculation of two-cluster system binding energy.

Thereby the intercluster potentials have been constructed on the basis of the definite assumptions, i.e. their parameters were fixed for the correct description of the elastic scattering processes phases and the bound states characteristics. In whole it is succeeded to reproduce the available experimental data for the scattering phases, $^{16}$N nucleus levels spectra and other characteristics of the bound states of this nucleus in n$^{15}$N-channel. In addition the suggested potentials of scattering and bound states satisfy the given above classification of the clusters orbital states according to the Young diagrams. Obtained potentials let further to consider the $E1$-transitions from the various $P$-states of n$^{15}$N-scattering to $^{16}$N nucleus ground state which is bound in n$^{15}$N-channel.

**4. Total cross sections**
Beginning the immediate consideration of the results for mentioned above $E1$-transitions to the ground state and the first three excited states of $^{16}$N nucleus let's note, that we succeeded to find the experimental data [37,38,39,40] for the total cross-section for n$^{15}$N-capture process only at three energy values 25, 152 and 370 keV [41], these results are presented in figures 2 by the black dots. The results of our calculation of the total cross-section of $^{3}P_2 \to {}^{3}D_2$ $E1$ capture process to the ground state are shown in figure 2 by dotted curve, the cross-sections for $\sigma(E1, {}^{3}P_1 \to {}^{3}D_2) + \sigma(E1, {}^{1}P_1 \to {}^{1}D_2)$ transitions also to the ground state are presented by dashed line, and their sum is given by solid curve. In these calculations the given above potentials (2)-(4) were used.

Due to the large width of the phase shift resonance, which is given by $P_1$-potential (2), the resonance width of the total cross-sections in this calculations is some overestimated, that leads to some overestimated value of the cross-section at energy 370 keV [41] which is close by resonance.

Then it may be noted, that if the parameters of the $P_1$-resonance potential are fixed over the phase resonance rather unambiguously, and for the bound state they are definitely selected on the basis of the description of its characteristics, so the parameters of the $^{3}P_2$-potential with the FS (3) leading to zero phases may take on other values. However, if one use, for example, more narrow potential with parameters 1000 MeV and 2.0 Fm$^{-2}$ which also leads to the near-zero phases, the results of the cross-sections calculations for the transition to the ground state differ on the value about 1%. This result demonstrates the weak influence of such potential geometry on the capture total cross-sections. Here only the near-zero value of scattering phase is important.

So the intercluster potential of the bound state constructed on the basis of quite evident requirements of the description of the binding energy, the root mean square radii of $^{16}$N nucleus and values of the AC in n$^{15}$N-channel let in whole to reproduce correctly the available experimental data for the radiative n$^{15}$N-capture total cross-sections at low energies [41]. In addition all used n$^{15}$N-potentials were constructed on the basis of the given above classification of the FS and AS according to the Young diagrams. However, it is difficult to draw some conclusions if there are available only three experimental points in the total cross-sections at 25÷370 keV [41]. So in the future it is desirable that the more detailed measurements of such cross-sections in the energy range from 1÷10 keV to 1.0÷1.2 MeV will be made. These measurements must totally define resonance width of this reaction at 921 keV [18] and the cross-section value at the resonance energy, that let to compare it with the results of these calculations which predict the cross-section in the resonance approximately equal to 180 μb.

Further clarify the effect of number of FS in the intercluster potentials to results of the total cross sections calculations. If we use the potential without the FS with the depth 5302.745 MeV and $\alpha$ = 20.0 Fm$^{-2}$ which leads to much the same width of resonance at 921 keV and doesn't conform with the given above classification of the forbidden states, so the results of calculation of the ground state capture total cross-sections of considered reaction are practically the same as the results shown by dashed curve obtained for potential (2). In addition, for the subsequent comparison of the results the ground state potential with one FS, i.e. incompatible with the given above classification, can be used. Let's consider, for example, the parameters

$$V_{\text{g.s.}} = 151.424599 \text{ MeV}, \quad \alpha_{\text{g.s.}} = 0.11 \text{ Fm}^{-2},$$

which lead to the binding energy 2.491000 MeV at the finite-difference method accuracy $10^{-6}$ MeV, the charge and mass radii 2.63 Fm and 2.77 Fm correspondingly and much the same asymptotic constant equal to 0.97. In this case the results of calculation of the sum total cross-sections for the ground state capture are practically the same as results shown by solid line for potentials (2-4).

Now let's notice, as the calculated cross-section is practically the straight line at the lowest energies from 1 eV to 10 keV (see solid curve in figure 2), then it may be approximated by the simple function of the form

$$\sigma_{ap}(\mu b) = 0.9907 \sqrt{E_n(\text{keV})} \ .$$

The given constant's value 0.9907 $\mu$b·keV$^{-1/2}$ has been defined over one point in the cross-sections at the minimal energy equal to 1 eV. Then it was found that the absolute value

$$M(E) = \left| [\sigma_{ap}(E) - \sigma_{theor}(E)] / \sigma_{theor}(E) \right|$$

of the relative departure of the calculated theoretical cross-section and approximation of this cross-section by the given above function at energies less than 10 keV is about 0.1%. If it is assumed, that this form of energy dependence of the total cross-section will be also conserved at lower energies, then one may give estimate of the cross-section value which, for example, at energy 1 meV ($10^{-3}$ eV = $10^{-6}$ keV) is equal to $10^{-3}$ $\mu$b.

## 5. Conclusion

As a result the total cross-sections for the considered n$^{15}$N-capture reaction at low energies weakly depend on the number of the FS in the interaction potentials. In other words if we use the phase-equivalent potentials of scattering and the bound states interaction, which are leading to the same quality of description of the basic characteristics of the bound states, notably, the asymptotic constant, then the results of the capture total cross-sections calculation practically don't depend on the FS number. This conclusion is opposite in whole to the observations made earlier for analysis of many other light nuclei in the cluster channels and capture reactions with them [6-11].

Considered here system and the capture process are the certain exceptions to the earlier observed strong dependence of the total cross-sections on the FS number in the definite partial potential, i.e. in the intercluster interaction for the given orbital momentum [6-8]. Usually the potentials with "incorrect" number of the FS led to much more difference between the cross-sections. In this case the FS number practically doesn't influence on the calculation results which now depend only on the resonance level width obtained on the basis of the used elastic $P_1$-scattering potential.

In conclusion let's note, that by this time there are 20 cluster systems considered by us on the basis of the potential cluster model with the orbital states classification according to the Young diagrams [6-11,17,23,24,42,43]. In these systems it is succeeded to obtain the quite reasonable results for the description of characteristics of the radiative nucleons or light clusters capture processes on atomic nuclei. In addition we always used the intercluster potentials consistent with the phases of the elastic scattering of clusters in the initial channel and characteristics of the bound states of nuclei generated in the reaction. Properties of these cluster nuclei and some characteristics of their bound states in the considerable cluster channels are presented in table 1.

**Acknowledgements**
In conclusion authors express their heartfelt gratitude to Dzhazairov-Kakhramanov A.V. and Burkova N.A. for the discussion of some issues touched upon in this work.

This work was supported by the Grant Program No. 0151/GF2 of the Ministry of Education and Science of the Republic of Kazakhstan: The study of thermonuclear processes in the primordial nucleosynthesis of the Universe.

**References**

[1] *Nuclear astrophysics* 1986 ed Barnes C A, Clayton D D *et al* (in Russian) (Moscow: Mir)
Barnes C A, Clayton D D and Schramm D N 1982 *Essays in Nuclear Astrophysics* Presented to William A. Fowler (Cambridge, UK: Cambridge University Press) p 562
[2] Adelberger E G *et al* 2011 *Rev. Mod. Phys.* **83** 195.
[3] Heil M *et al* 1998 *Astrophys. J.* **507** 997
Guimaraes V and Bertulani C A 2009 arXiv:0912.0221v1 [nucl-th] 1 Dec 2009
Masayuki Igashira and Toshiro Ohsaki 2004 *Sci. Tech. Adv. Materials* **5** 567
Nagai Y *et al* 1996 *Hyperfine Interactions* **103** 43
Liu Z H *et al* 2001 *Phys. Rev. C* **64** 034312
Horvath A *et al* 2002 *Astrophys. J.* **570** 926
[4] Kapitonov I M, Ishkhanov B S and Tutyn' I A 2009 *Nucleosynthesis in the Universe* (in Russian) (Moscow: Librokom) available online at http://nuclphys.sinp.msu.ru/nuclsynt.html
[5] Dubovichenko S B 2011 *Selected methods of nuclear astrophysics* Series *Kazakhstan space research* V 9 (Almaty: Fessenkov V G Astrophysical Institute) arXiv:1201.3003v2 [nucl-th] available online at http://arxiv.org/abs/1201.3003.
[6] Dubovichenko S B 2012 *Selected methods of nuclear astrophysics* Second edition, corrected and expanded (Saarbrucken, Germany: Lambert Acad. Publ. GmbH&Co. KG) available online at https://www.lap-publishing.com/catalog/details/store/gb/book/978-3-8465-8905-2/Избранные-методы-ядерной-астрофизики.
[7] Dubovichenko S B 2011 *Thermonuclear processes of the Universe* Second edition, corrected and expanded Series *Kazakhstan space research* V 7 (Almaty: A-tri) arXiv:1012.0877v3 [nucl-th] available online at http://arxiv.org/abs/1012.0877.
[8] Dubovichenko S B 2012 *Thermonuclear Processes of the Universe* (New-York: NOVA Sci. Publ.) available online at
https://www.novapublishers.com/catalog/product_info.php?products_id=31125
[9] Dubovichenko S B and Uzikov Yu N 2011 *Phys. Part. Nucl.* **42** 251 available online at http://www.springerlink.com/content/f4681n1654732860/
[10] Dubovichenko S B and Dzhazairov-Kakhramanov A V 2012 *The Big Bang: Theory, Assumptions and Problems* (New-York: NOVA Sci. Publ.) available online at
https://www.novapublishers.com/catalog/product_info.php?products_id=21109.
[11] Dubovichenko S B and Dzhazairov-Kakhramanov A V 2012 *Int. Jour. Mod. Phys.* **21** 1250039
[12] Neudatchin V G, Sakharuk A A and Smirnov Yu F 1992 *Sov. J. Part. Nucl.* **23** 210
Neudatchin V G *et al* 1992 *Phys. Rev. C* **45** 1512
[13] Nemez O F *et al* 1988 *Nucleon associations in the atomic nuclei and nuclear reactions of multinucleon transfers* (in Russian) (Kiev: Naukova dumka)
[14] Neudatchin V G and Smirnov Yu F 1969 *Nucleon associations in light nuclei* (in Russian) (Moscow: Nauka)
[15] Dubovichenko S B 1998 *Phys. Atom. Nucl.* **61** 162; 1995 *Phys. Atom. Nucl.* **58** 1174; 1995 *Phys. Atom. Nucl.* **58** 1295; 1995 *Phys. Atom. Nucl.* **58** 1866; Dubovichenko S B and Dzhazairov-Kakhramanov A V 1995 *Phys. Atom. Nucl.* **58** 579; 1994 *Phys. Atom. Nucl.* **57** 733; 1990 *Soviet J. Nucl. Phys.* **51** 971
[16] Dubovichenko S B 2004 *Properties of the light nuclei in the potential cluster model* (Almaty: Daneker) arXiv:1006.4944v2 [nucl-th] available online at http://arxiv.org/abs/1006.4944

[17] Dubovichenko S B 2013 *Rus. Phys. J.* **56** (In press)
[18] Tilley D R, Weller H R and Cheves C M 1993 *Nucl. Phys. A* **564** 1
[19] Ajzenberg-Selove F 1991 *Nucl. Phys. A* **523** 1
[20] Itzykson C and Nauenberg M 1966 *Rev. Mod. Phys.* **38** 95
[21] Dubovichenko S B and Dzhazairov-Kakhramanov A V 2009 *Eur. Phys. J. A* **39** 139
[22] Dubovichenko S B 2011 *Rus. Phys. J.* **54** 157
Dubovichenko S B 2012 *Rus. Phys. J.* **55** 138
http://www.springerlink.com/content/?Author=S.+B.+Dubovichenko
[23] Dubovichenko S B, Dzhazairov-Kakhramanov A V and Burkova N A 2012 *Jour. Nucl. Part. Phys.* **2** 6
Dubovichenko S B 2012 *Phys. Atom. Nucl.* **75** 173
[24] Dubovichenko S B and Dzhazairov-Kakhramanov A V 2012 arXiv:1201.1741v7 [nucl-th] 18 Jul. 2012 available online at http://xxx.lanl.gov/abs/1201.1741
Dubovichenko S B and Burkova N A 2012 arXiv:1202.1420v2 [nucl-th] 16 Apr. 2012 available online at http://arxiv.org/abs/1202.1420.
[25] Angulo C *et al* 1999 *Nucl. Phys. A* **656** 3
[26] Dubovichenko S B and Dzhazairov-Kakhramanov A V 1997 *Phys. Part. Nucl.* **28** 615 available online at http://www1.jinr.ru/Archive/Pepan/1997-v28/v-28-6/4.htm.
[27] Dubovichenko S B 2006 *Calculation methods of nuclear characteristics: Methods, calculations and programs* (Almaty: Complex) arXiv:1006.4947v2 [nucl-th] available online at http://arxiv.org/abs/1006.4947
[28] Dubovichenko S B 2012 *Calculation methods of nuclear characteristics: Nuclear and thermonuclear processes*. 2$^{nd}$ edition (Saarbrucken, Germany: Lambert Acad. Publ. GmbH&Co. KG) available online at https://www.lap-publishing.com/catalog/details//store/ru/book/978-3-659-21137-9/методы-расчета-ядерных-характеристик
[29] Kukulin V I *et al* 1984 *Nucl. Phys. A* **417** 128
Kukulin V I *et al* 1986 *Nucl. Phys. A* **453** 365
Kukulin V I *et al* 1990 *Nucl. Phys. A* **517** 221
[30] http://physics.nist.gov/cgi-bin/cuu/Value?mud|search_for=atomnuc! .
[31] http://cdfe.sinp.msu.ru/services/ground/NuclChart_release.html .
[32] http://physics.nist.gov/cuu/Constants/ .
[33] Dubovichenko S B and Dzhazairov-Kakhramanov A V 2012 *Ann. der Phys.* **524** 850
[34] Burkova N A, Zhaksybekova K A and Zhusupov M A 2009 One-nucleon spectroscopy of light nuclei *Phys. Part. Nucl.* **40** 162-205
[35] Plattner G R and Viollier R D 1981 *Nucl. Phys. A* **365** 8
[36] Huang J T, Bertulani C A and Guimaraes V 2010 *Atom. Data and Nucl. Data Tabl.* **96** 824 arXiv:0810.3867v2 [nucl-th], 19 May 2009.
[37] http://cdfe.sinp.msu.ru/exfor/index.php .
[38] http://www.nndc.bnl.gov/exfor/exfor00.htm .
[39] http://xxx.lanl.gov/find/nucl-ex .
[40] Herndl H *et al* 1999 *Phys. Rev. C* **60** 064614 nucl-th/9908087v1 31 Aug 1999
[41] Meissner J *et al* 1996 *Phys. Rev. C* **53** 977
[42] Dubovichenko S B 2013 *Phys. Atom. Nucl.* **76** (In press)
[43] Dubovichenko S B and Dzhazairov-Kakhramanov A V 2011 *Bull. Russ. Acad. of Sci. Ser. Phys.* **75** 1517

**TITLES OF FIGURES**

Figure 1. Phases of the elastic n$^{15}$N-scattering in $P_1$-wave. Curves are obtained for the different potentials described in the text.

Figure 2. Total cross-sections of the radiative n$^{15}$N-capture. Experimental data: ● – [41]. Curves correspond to the calculation of the total cross-sections for the transitions to the ground state in the region 1 eV – 1.0 MeV.

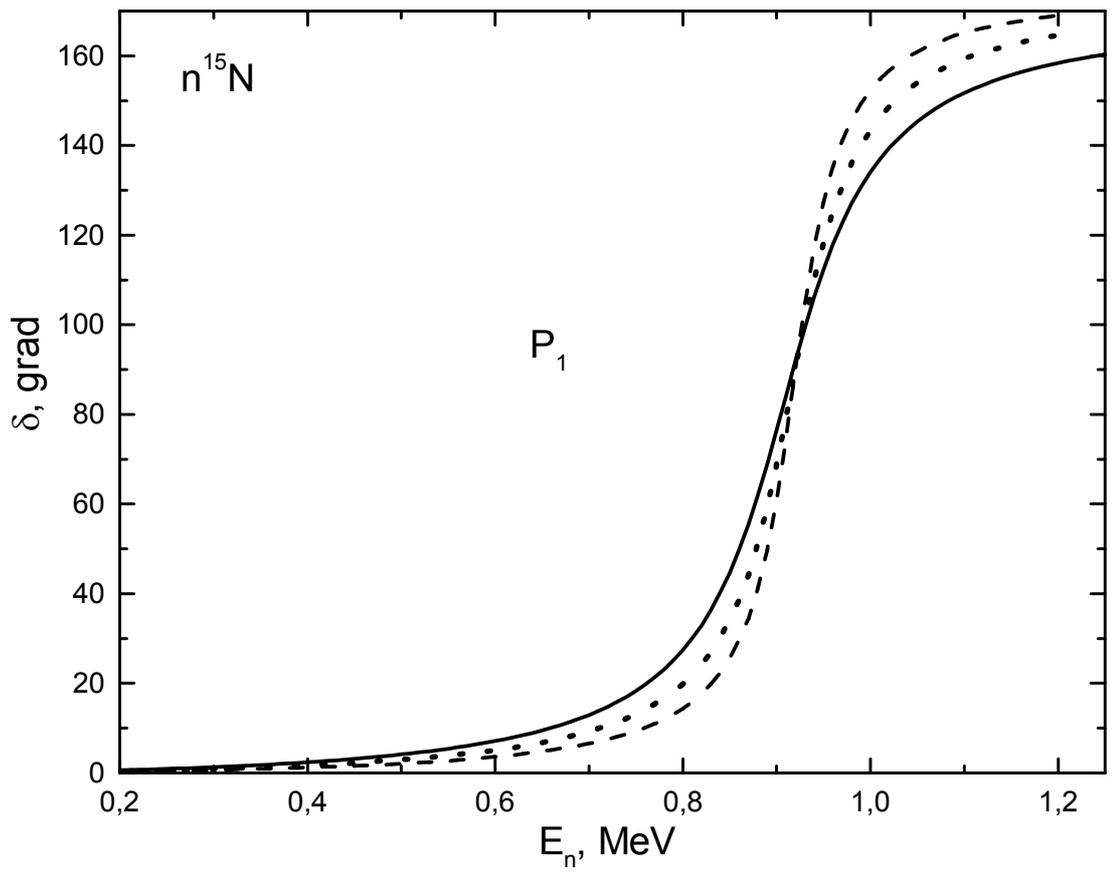

Fig.1

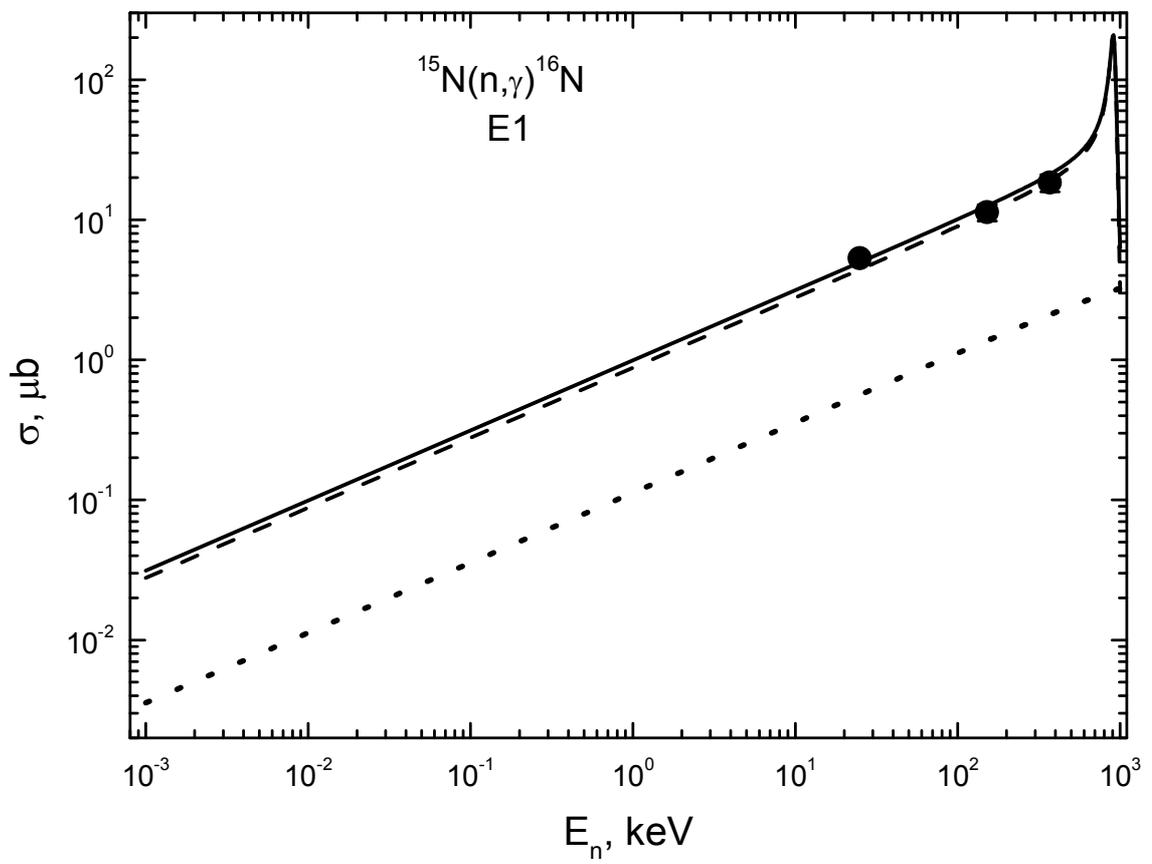

Fig.2

**Table 1.** Nuclei and cluster systems characteristics and our references in which the radiative capture processes with these particles were considered.

| № | Nucleus ($J^\pi, T$) | Cluster channel | $T_z$ | $T$ | Reference |
|---|---|---|---|---|---|
| 1 | $^3$He ($1/2^+, 1/2$) | p$^2$H | +1/2 + 0 = +1/2 | 1/2 | [7]] |
| 2 | $^3$H ($1/2^+, 1/2$) | n$^2$H | -1/2 + 0 = -1/2 | 1/2 | [6,22] |
| 3 | $^4$He ($0^+, 0$) | p$^3$H | +1/2 - 1/2 = 0 | 0 + 1 | [7-11] |
| 4 | $^6$Li ($1^+, 0$) | $^2$H$^4$He | 0 + 0 = 0 | 0 | [7-11] |
| 5 | $^7$Li ($3/2^-, 1/2$) | $^3$H$^4$He | -1/2 + 0 = -1/2 | 1/2 | [7-11] |
| 6 | $^7$Be ($3/2^-, 1/2$) | $^3$He$^4$He | +1/2 + 0 = +1/2 | 1/2 | [7-11] |
| 7 | $^7$Be ($3/2^-, 1/2$) | p$^6$Li | +1/2 + 0 = +1/2 | 1/2 | [7-11] |
| 8 | $^7$Li ($3/2^-, 1/2$) | n$^6$Li | -1/2 + 0 = -1/2 | 1/2 | [6] |
| 9 | $^8$Be ($0^+, 0$) | p$^7$Li | +1/2 - 1/2 = 0 | 0 + 1 | [7-11] |
| 10 | $^8$Li ($2^+, 1$) | n$^7$Li | -1/2 - 1/2 = -1 | 1 | [6,33] |
| 11 | $^{10}$B ($3^+, 0$) | p$^9$Be | +1/2 - 1/2 = 0 | 0 + 1 | [7-11] |
| 12 | $^{10}$Be ($0^+, 1$) | n$^9$Be | -1/2 - 1/2 = -1 | 1 | [6] |
| 13 | $^{13}$N ($1/2^-, 1/2$) | p$^{12}$C | +1/2 + 0 = +1/2 | 1/2 | [7-11] |
| 14 | $^{13}$C ($1/2^-, 1/2$) | n$^{12}$C | -1/2 + 0 = -1/2 | 1/2 | [6,24] |
| 15 | $^{14}$N ($1^+, 0$) | p$^{13}$C | +1/2 - 1/2 = 0 | 0 + 1 | [23] |
| 16 | $^{14}$C ($0^+, 1$) | n$^{13}$C | -1/2 – ½ = -1 | 1 | [6,24] |
| 17 | $^{15}$C ($1/2^+, 3/2$) | n$^{14}$C | -1/2 – 1 = -3/2 | 3/2 | [6,42] |
| 18 | $^{15}$N ($1/2^-, 1/2$) | n$^{14}$N | -1/2 + 0 = -1/2 | 1/2 | [17,42] |
| 19 | $^{16}$N ($2^-, 1$) | n$^{15}$N | -1/2 +-1/2 = -1 | 1 | This article |
| 20 | $^{16}$O ($0+, 0$) | $^4$He$^{12}$C | 0 + 0 = 0 | 0 | [7-11,43] |